%% file: MF1217rv.tex
\documentclass[useAMS,usenatbib]{mn2e}

\input{psfig}
\input{macros_yun}

\begin{document}


\title[On the Assembly History of Dark Matter Haloes]
      {On the Assembly History of Dark Matter Haloes}
\author[Li, Mo, van den Bosch \& Lin]
       {Yun~Li$^1$, H.J.~Mo$^1$, Frank C. van den Bosch$^{2}$, W.P. Lin$^{3}$
       \thanks{E-mail: liyun@nova.astro.umass.edu}\\
        $^1$Department of Astronomy, University of Massachusetts, 
            MA 01003, USA\\
        $^2$Max-Planck-Institute for Astronomy, K\"onigstuhl 17, 
            D-69117 Heidelberg, Germany \\
        $^3$Shanghai Astronomical Observatory, 80 Nandan Rd., 
            Shanghai 200030, China}


\date{}

\pagerange{\pageref{firstpage}--\pageref{lastpage}}
\pubyear{2000}

\maketitle

\label{firstpage}


\begin{abstract}
  We study the mass  assembly history (MAH)  of dark matter haloes. We
  compare  MAHs obtained using (i) merger   trees constructed with the
  extended  Press    Schechter     (EPS)  formalism,   (ii)  numerical
  simulations, and (iii) the  Lagrangian perturbation code  PINOCCHIO. 
  We show   that the PINOCCHIO MAHs  are  in  excellent agreement with
  those obtained using numerical  simulations, while the EPS formalism
  predicts MAHs that  occur too late.   PINOCCHIO, which is much  less
  CPU  intensive than N-body  simulation,   can be  run on  a simple
  personal  computer,    and does not   require   any labor intensive
  post-simulation analysis, therefore provides  a unique and  powerful
  tool to investigate the growth history of dark matter haloes.  Using
  a suite of 55 PINOCCHIO simulations, with $2563$ particles each, we
  study the MAHs of 12,924  cold dark matter  haloes in a $\Lambda$CDM
  concordance  cosmology.  This is  by far  the  largest set of haloes
  used for any  such analysis.  For each MAH  we derive four different
  formation redshifts, which characterize  different epochs during the
  assembly  history of a dark  matter halo.  We  show that haloes less
  massive than  the  characteristic non-linear  mass  scale  establish
  their potential wells  much before they acquire  most of their mass. 
  The time   when a halo reaches  its  maximum virial velocity roughly
  divides  its mass assembly into  two phases,  a fast accretion phase
  which is dominated  by  major mergers, and  a  slow  accretion phase
  dominated by minor mergers.  Each  halo experiences about $3 \pm  2$
  major mergers  since  its main progenitor had   a mass equal to  one
  percent of  the final halo   mass.  This major merger  statistic  is
  found to be   virtually  independent  of halo  mass.   However,  the
  average  redshift at  which these  major  mergers occur is strongly
  mass dependent, with  more massive  haloes experiencing their  major
  mergers later.   
\end{abstract}

\begin{keywords}
cosmology: theory ---
galaxies: formation ---
galaxies: haloes ---
dark matter.
\end{keywords}

\section{Introduction}
\label{sec:intro}

The cold dark matter (CDM)  paradigm has become the standard framework
for  the  formation  of  large-scale structure  and  galaxies.   Small
fluctuations  in   the  initial  density   field  grow  by   means  of
gravitational instability until they  collapse to form virialized dark
matter haloes. This  growth process is hierarchical in  the sense that
small clumps  virialize first, and aggregate  successively into larger
and larger objects.   Galaxies form from the gas  that is shock heated
by the  gravitational collapse and  then subsequently cools  (White \&
Rees 1978; but see also Birnboim  \& Dekel 2003 and Keres \etal 2004). 
Therefore,  a proper understanding  of galaxy  formation relies  on an
accurate  description of  the structure  and assembly  of  these dark
matter  haloes. This  problem  is  tackled by  a  combination of  both
N-body  simulations   and  analytical  models.    Although  N-body
simulations have the advantage that  they follow the formation of dark
matter haloes into the non-linear  regime, they are expensive, both in
terms of labor (analyzing  the simulations) and CPU time.  Therefore,
accurate analytical  models are always useful.  The  most developed of
these  is the  Press-Schechter  (PS) formalism,  which  allows one  to
compute  the (unconditional)  halo mass  function (Press  \& Schechter
1974).   Bond \etal  (1991), Bower  (1991)  and Lacey  \& Cole  (1993)
extended  the  PS formalism,  using  the  excursion  set approach,  to
compute conditional  mass functions.  These allow  the construction of
merger  histories,  the  computation  of  halo  formation  times,  and
detailed  studies of spatial  clustering and  large scale  bias (e.g.,
Kauffmann \&  White 1993; Mo  \& White 1996,  2002; Mo, Jing  \& White
1996,  1997; Catelan  \etal 1998;  Sheth 1998;  Nusser \&  Sheth 1999;
Somerville \& Kolatt 1999; Cohn, Bagla \& White 2001).

Numerous studies in  the past have tested the  predictions of extended
Press-Schechter (EPS)  theory against numerical  simulations. Although
the  unconditional  mass  function  was  found  to  be  in  reasonable
agreement, it  systematically over (under) predicts the  number of low
(high)  mass haloes  (e.g., Jain  \& Bertschinger  1994;  Tormen 1998;
Gross \etal 1998; Governato  \etal 1999; Jenkins \etal 2001).  Similar
discrepancies have been found  regarding the conditional mass function
(Sheth  \&  Lemson 1999;  Somerville  \etal  2000),  which results  in
systematic offsets of the halo formation times predicted by EPS (e.g.,
van den Bosch 2002a).  Finally,  Bond \etal (1991) have shown that the
PS  approach achieves  a very  poor agreement  on  an object-by-object
basis when compared with simulations (for a review, see Monaco 1998).

It  is generally  understood that  these discrepancies  stem  from the
assumption of  spherical collapse. Numerous  studies have investigated
schemes to improve the EPS formalism by using ellipsoidal, rather than
spherical collapse  conditions, thereby  taking proper account  of the
aspherical nature of collapse in  a CDM density field (e.g., Sheth, Mo
\& Tormen 2001,  hereafter SMT01; Sheth \& Tormen  2002; Chiueh \& Lee
2001;   Lin,  Chuieh  \&   Lee  2002).    Although  this   results  in
unconditional mass  functions that are  in much better  agreement with
numerical  simulations (e.g.,  SMT01; Jenkins  \etal 2001),  they have
been unable thus far to yield conditional mass functions of sufficient
accuracy that reliable merger trees can be constructed.

Despite its systematic errors  and uncertainties, the PS formalism has
remained  the   standard  analytical  approach    in galaxy  formation
modeling.  In particular, the extended Press-Schechter theory is used
extensively  to compute merger   histories and mass assembly histories
(hereafter MAHs)  which serve  as  the back-bone for models  of galaxy
formation  (e.g., Kauffmann, White  \&  Guiderdoni 1993; Somerville \&
Primack  1999; Cole  \etal   2000;  van  den  Bosch  2001; Firmani  \&
Avila-Reese 2000).    This  may have  profound   implications for  the
accuracy  of these models.  For instance,  the mass assembly histories
of dark matter haloes  are  expected to  impact on the  star formation
histories of the galaxies that form  inside these haloes. In addition,
the merger and mass assembly history of  individual haloes may also be
tightly related to their  internal  structure.  As shown by   Wechsler
\etal (2002;  hereafter W02)   and  Zhao \etal (2003a;b), the   MAH is
directly related to the  concentration  of the resulting dark   matter
halo (see also Navarro, Frenk \&  White 1997; Bullock \etal 2001; Eke,
Navarro  \& Steinmetz 2001). Errors in  the mass assembly histories of
dark matter haloes    may therefore result in erroneous    predictions
regarding   the star formation history  and  the rotation curve shapes
and/or  the zero-point  of    the Tully-Fisher relation (e.g.,   Alam,
Bullock \& Weinberg 2002; Zentner  \& Bullock 2002;  Mo \& Mao (2000);
van den Bosch, Mo \& Yang 2003).  Clearly, a detailed understanding of
galaxy formation requires a description  of the growth history of dark
matter haloes that is    more  accurate than EPS. Although    $N$-body
simulations are probably the most reliable means of obtaining accurate
assembly histories of dark matter haloes, they are computationally too
expensive for some purposes.

As  an alternative  to  the EPS  formalism  and N-body  simulations,
perturbative techniques  have been developed that  describe the growth
of dark  matter haloes  in a given  numerical realization of  a linear
density field. These include, amongst others, the truncated Zel'dovich
(1970) approximation   (Borgani,  Coles   \&  Moscardini   1994),  the
peak-patch  algorithm (Bond  \& Myers  1996a,b) and  the  merging cell
model (Rodriguez \&  Thomas 1996; Lanzoni, Mamon \&  Guiderdoni 2000). 
Recently, Monaco, Theuns \& Taffoni (2002b) developed a numerical code
that  uses local  ellipsoidal collapse  approximations (Bond  \& Myers
1996a;  Monaco  1995)  within  Lagrangian  Perturbation  Theory  (LPT,
Buchert \&  Ehlers 1993; Catelan  1995).  This code,  called PINOCCHIO
(PINpointing Orbit-Crossing Collapsed  HIerarchical Objects), has been
shown  to   yield  accurate  mass  functions,   both  conditional  and
unconditional (Monaco \etal 2002a,b;  Taffoni, Monaco \& Theuns 2002),
and  is therefore ideally  suited to  study halo  assembly histories,
without   having  to  rely   on  computationally   expensive  N-body
simulations.  

This  paper is  organized as  follows. In  Section~\ref{sec:theory} we
give  a  detailed   overview  of  (extended)  Press-Schechter  theory,
including  a discussion  of  its short-comings  and its  modifications
under  ellipsoidal collapse  conditions, and  describe  the Lagrangian
perturbation code  PINOCCHIO. In Section~\ref{sec:sim}  we compare the
MAHs  obtained   from  PINOCCHIO,   the  EPS  formalism,   and  N-body
simulations.  We show that PINOCCHIO yields MAHs that are in excellent
agreement  with numerical  simulations,  and do  not  suffer from  the
shortcomings of the EPS formalism. In the second part of this paper we
then  analyze  a  large,  statistical  sample of  MAHs  obtained  with
PINOCCHIO   for  haloes  spanning   a  wide   range  in   masses.   In
Section~\ref{sec:ftime} we  use these MAHs to study,  in a statistical
sense, various  characteristic epochs and events in  the mass assembly
history of  a typical  CDM halo.  We  analyze the statistics  of major
merger events in Section~\ref{sec:majmerprop}. Finally, 
Section~\ref{sec:concl} summarizes our results.

\section{Theoretical background}
\label{sec:theory}

\subsection{Extended Press-Schechter theory}
\label{sec:EPS}

In  the standard  model for  structure formation  the  initial density
contrast  $\delta({\bf   x})  =  \rho({\bf  x})/\bar{\rho}   -  1$  is
considered  to  be  a   Gaussian  random  field,  which  is  therefore
completely specified by the power spectrum $P(k)$.  As long as $\delta
\ll  1$ the  growth of  the perturbations  is linear  and $\delta({\bf
  x},t_2) =  \delta({\bf x},t_1)  D(t_2)/D(t_1)$, where $D(t)$  is the
linear growth factor linearly  extrapolated to the present time.  Once
$\delta({\bf x})$ exceeds a critical threshold $\delta^{0}_{\rm crit}$
the  perturbation  starts to  collapse  to  form  a virialized  object
(halo).   In the  case  of spherical  collapse $\delta^{0}_{\rm  crit}
\simeq 1.68$.   In what  follows we define  $\delta_0$ as  the initial
density contrast field linearly  extrapolated to the present time.  In
terms of  $\delta_0$, regions that  have collapsed to  form virialized
objects at  redshift $z$  are then associated  with those  regions for
which $\delta_0 > \delta_c(z) \equiv \delta^{0}_{\rm crit}/D(z)$.

In order to assign masses to these collapsed regions, the PS formalism
considers  the density  contrast  $\delta_0$ smoothed  with a  spatial
window function  (filter) $W(r;R_f)$.  Here $R_f$  is a characteristic
size of the filter, which is used to compute a halo mass $M = \gamma_f
\bar{\rho} R_f3/3$,  with $\bar{\rho}$ the  mean mass density  of the
Universe  and $\gamma_f$  a  geometrical factor  that  depends on  the
particular choice of  filter. The {\it ansatz} of  the PS formalism is
that the fraction of mass that  at redshift $z$ is contained in haloes
with masses  greater than  $M$ is equal  to two times  the probability
that   the   density  contrast   smoothed   with  $W(r;R_f)$   exceeds
$\delta_c(z)$.  This  results in the  well known PS mass  function for
the comoving number density of haloes:
\begin{eqnarray}
\label{PS}
\lefteqn{{{\dd}n \over {\dd} \, {\rm ln} \, M}(M,z) \, {\dd}M =}
\nonumber \\ & & \sqrt{2 \over \pi} \, \bar{\rho} \, {\delta_c(z)
\over \sigma2(M)} \, \left| {{\dd} \sigma \over {\dd} M}\right| \,
{\rm exp}\left[-{\delta_c2(z) \over 2 \sigma2(M)}\right] \, {\dd}M
\end{eqnarray}
(Press \& Schechter 1974). Here  $\sigma2(M)$ is the mass variance of
the smoothed density field given by
\begin{equation}
\label{variance}
\sigma2(M) = {1 \over 2 \pi2} \int_{0}^{\infty} P(k) \;
\widehat{W}2(k;R_f) \; k2 \; {\dd}k.
\end{equation}
with $\widehat{W}(k;R_f)$  the Fourier transform  of $W(r;R_f)$.

The {\it extended} Press-Schechter (EPS) model developed by Bond \etal
(1991), is based  on the excursion set  formalism.  For each point one
constructs `trajectories' $\delta(M)$  of the  linear density contrast
at that   position as function  of  the smoothing   mass $M$.  In what
follows  we  adopt the notation  of Lacey  \& Cole  (1993) and use the
variables $S = \sigma2(M)$ and  $\omega = \delta_c(z)$ to label  mass
and redshift, respectively.  In the limit $R_f \rightarrow \infty$ one
has that $S  = \delta(S) = 0$,  which  can be  considered the starting
point of the  trajectories.  Increasing $S$ corresponds to  decreasing
the filter mass $M$, and $\delta(S)$  starts to wander away from zero,
executing a random walk  (if the filter is  a sharp $k$-space filter). 
The fraction of matter in collapsed objects  in the mass interval $M$,
$M+{\rm d}M$  at redshift $z$ is now  associated with  the fraction of
trajectories that have their    {\it  first upcrossing}  through   the
barrier   $\omega = \delta_c(z)$ in  the   interval $S$, $S+{\rm d}S$,
which is given by
\begin{equation}
\label{probS}
P(S ,\omega) \; {\dd}S = {1  \over \sqrt{2 \pi}} \; 
{\omega  \over S^{3/2}} \; 
{\rm exp}\left[-{\omega2 \over 2 S}\right] \; {\dd}S
\end{equation}
(Bond \etal 1991;  Bower 1991; Lacey \& Cole  1993).  After conversion
to  number counting,  this  probability function  yields  the PS  mass
function  of equation~(\ref{PS}).  Note  that this  approach does  not
suffer  from  the  arbitrary  factor  two in  the  original  Press  \&
Schechter approach.

Since for random walks the upcrossing probabilities are independent of
the path  taken    (i.e., the upcrossing  is  a   Markov process), the
probability for a change $\dS$ in a time step $\dW$ is simply given by
equation~(\ref{probS}) with $S$  and $\omega$ replaced with $\dS$  and
$\dW$, respectively. This allows one to  immediate write down the {\it
conditional}  probability that a  particle in a halo  of mass $M_2$ at
$z_2$ was embedded in a halo of mass $M_1$ at $z_1$ (with $z_1 > z_2$)
as
\begin{eqnarray}
\label{probSS}
\lefteqn{P(S_1,\omega_1 \vert  S_2,\omega_2) \; {\dd}S_1 =} \nonumber \\
& & {1  \over \sqrt{2 \pi}} \;
{(\omega_1    -   \omega_2)    \over   (S_1    -    S_2)^{3/2}} \; {\rm
exp}\left[-{(\omega_1 - \omega_2)2 \over 2 (S_1 - S_2)}\right] \;
{\dd}S_1 
\end{eqnarray}
Converting from  mass weighting to  number weighting, one  obtains the
average number  of progenitors  at $z_1$ in  the mass  interval $M_1$,
$M_1 + {\rm d}M_1$ which by  redshift $z_2$ have merged to form a halo
of mass $M_2$:
\begin{eqnarray}
\label{condprobM}
\lefteqn{{{\dd}N \over {\dd}M_1}(M_1,z_1 \vert M_2,z_2) \; {\dd}M_1 =} 
\nonumber \\
& & {M_2 \over
M_1} \; P(S_1,\omega_1 \vert S_2,\omega_2) \; 
\left\vert {{\dd}S \over {\dd M}} \right\vert \; {\dd}M_1.
\end{eqnarray}
This  conditional  mass  function  can be  combined  with  Monte-Carlo
techniques to construct merger histories (also called merger trees) of
dark matter haloes.

\subsection{Ellipsoidal collapse}
\label{sec:ellips}

In an attempt to improve the inconsistencies between EPS and numerical
simulations  (see   Section~\ref{sec:intro}),   various authors   have
modified the EPS formalism  by   considering ellipsoidal rather   than
spherical  collapse.   For  ellipsoidal  density   perturbations,  the
conditions for collapse not  only depend on   the self-gravity of  the
perturbation, but also  on the tidal  coupling with  the external mass
distribution; external shear can actually  rip overdensities apart and
thus prevent them from collapsing.   Since smaller mass  perturbations
typically  experience a  stronger shear  field,  they tend  to be more
ellipsoidal.  Therefore, it is to be expected  that the assumptions of
spherical collapse in the standard EPS formalism are more accurate for
more massive haloes, whereas modifications associated with ellipsoidal
collapse will be  more dramatic  for smaller mass  haloes.  The way in
which  ellipsoidal collapse  modifies  the  halo formation times  with
respect to the EPS predictions depends  on the definition of collapse. 
Ellipsoidal   perturbations  collapse  independently  along  the three
different directions  defined by the eigen  vectors of the deformation
tensor  (defined as the second derivative  of the linear gravitational
potential). It is customary to  associate the first axis collapse with
the formation  of a 2-dimensional   pancake-like structure, the second
axis collapse with the formation of a  1-dimensional filament, and the
third axis  collapse  with the formation of  a  dark matter halo. Most
authors indeed have associated halo formation with the collapse of the
third axis (e.g., Bond \& Myers 1996a;  Audit, Teyssier \& Alimi 1997;
Lee \&  Shandarin 1998; SMT01), though some  have considered the first
axis collapse instead (e.g., Bertschinger  \& Jain 1994; Monaco 1995). 
For first-axis collapse one predicts that haloes  form earlier than in
the   spherical case, whereas   the opposite  applies when considering
third-axis    collapse.  Clearly,  the   implications   of considering
ellipsoidal rather than  spherical collapse depend  sensitively on the
collapse definition.

In order  to incorporate ellipsoidal collapse in  a PS-like formalism,
one  needs to  obtain  an  estimate of  the  critical overdensity  for
collapse $\delta_{ec}$.  Various studies  have attempted such schemes. 
For instance, SMT01 used the ellipsoidal collapse model to obtain
\begin{equation}
\label{ellips}
\delta_{ec}(M,z) = \delta_{c}(z) \left( 1 + 0.47 \left[{\sigma2(M)
\over \delta2_{c}(z)} \right]^{0.615}\right).
\end{equation}
Here $\delta_c(z)$  is the standard  value for the  spherical collapse
model.   Solving for  the upcrossing  statistics with  this particular
barrier shape  results in  halo mass functions  that are  in excellent
agreement  with those  found  in simulations  (Sheth  \& Tormen  1999;
Jenkins \etal  2001). Unfortunately, no analytical  expression for the
conditional  mass function  is  known for  a  barrier of  the form  of
equation~(\ref{ellips}), and  one has to resort  to either approximate
fitting  functions  (Sheth  \&  Tormen   2002),  or  one  has  to  use
time-consuming  Monte-Carlo simulations  to  determine the  upcrossing
statistics  (Chiueh  \&  Lee  2001;  Lin \etal  2002).   Although  the
resulting conditional mass  functions ${{\dd}N \over {\dd}M_1}(M_1,z_1
\vert M_2,z_2)  \; {\dd}M_1$ have been  found to be  in good agreement
with  numerical simulations if  a relatively  large look-back  time is
considered (i.e., if $\Delta z =  z_2-z_1 \gta 0.5$), there is still a
large disagreement for  small $\Delta z$. This is  probably due to the
neglect of  correlations between scales in the  excursion set approach
(Peacock \& Heavens  1990; Sheth \& Tormen 2002).  This is unfortunate
as it does not allow these  methods to be used for the construction of
merger histories or  MAHs.  Lin \etal (2002) tried  to circumvent this
problem  by introducing  a  small  mass gap  between  parent halo  and
progenitor halo,  i.e., each  time step they  require that $S_1  - S_2
\geq  f  \, \delta_c2(z_2)$.   Upon  testing  their conditional  mass
function with  this mass gap  against numerical simulations  they find
good agreement for  $f = 0.06$, and claim  that with this modification
the  excursion set  approach {\it  can}  be used  to construct  merger
histories under  ellipsoidal collapse conditions.   However, they only
tested  their conditional  mass  functions for  $\Delta  z \geq  0.2$,
whereas accurate  merger histories require  significantly smaller time
steps.  For instance,  van den Bosch (2002a) has  argued for timesteps
not  larger than  $\Delta \omega  = \omega_1  - \omega_2  \simeq 0.1$,
which,  for  an Einstein-de  Sitter  (EdS)  cosmology, corresponds  to
$\Delta z  \simeq 0.06$ (see  also discussion in Somerville  \& Kolatt
1999).  Furthermore, with the mass  gap suggested by Lin \etal (2002),
each time step there is a minimum amount of mass accreted by the halo,
which  follows  from  $S_1  -  S_2  =  f  \,  \delta_c2(z_2)$.   This
introduces a  distinct maximum to  the halo half-mass  formation time,
the value of which depends sensitively on the actual time-steps taken.
To test  this, we constructed MAHs  of CDM haloes using  the method of
van  den  Bosch  (2002a)  but  adopting  the  conditional  probability
function of Lin  \etal (2002). This resulted in MAHs  that are in very
poor agreement with numerical  simulations. In particular, the results
were found to depend strongly on the value of $\Delta \omega$ adopted.

In  summary, although introducing  ellipsoidal collapse  conditions in
the excursion  set formalism has allowed the  construction of accurate
unconditional mass functions, there  still is no reliable method based
on the EPS  formalism that allows the construction  of accurate merger
histories and/or MAHs.
\begin{figure*}
\centerline{\psfig{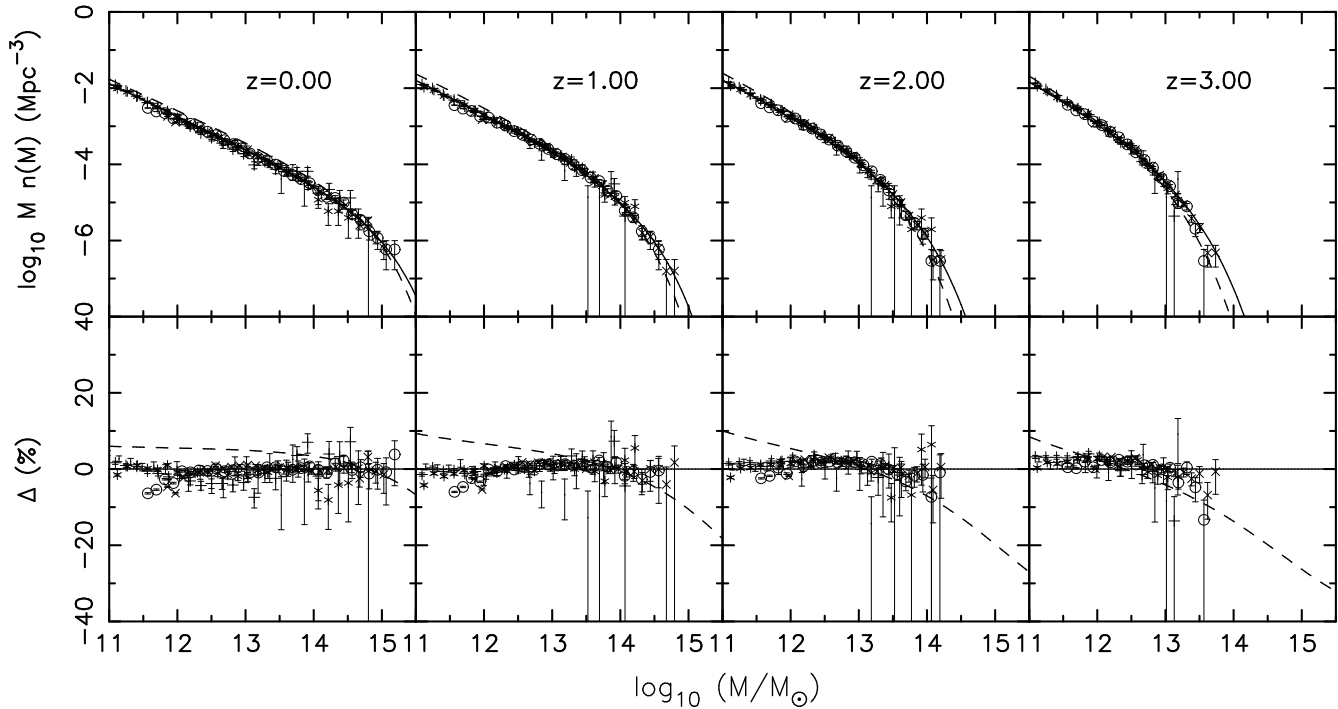}}
\caption{Panels in the upper row show the (unconditional) halo mass  
  functions at 4 different  redshifts, as indicated. Different symbols
  (each    with Poissonian error    bars)   correspond to  5 different
  PINOCCHIO simulations randomly selected from P0, 
  each with a different mass resolution. Dashed
  and solid lines  correspond  to  the PS  and SMT01  mass  functions,
  respectively, and are shown for comparison.  Panels in the lower row
  show  the  percentual difference    between  the PS and  SMT01  mass
  functions  (dashed  lines) and that between  the  PINOCCHIO  and the
  SMT01 mass functions (symbols with errorbars).  Clearly, the PS mass
  function overestimates (underestimates)  the number of  small (high)
  mass  haloes, while PINOCCHIO  yields   mass functions that  are  in
  excellent agreement  with  SMT01    (and  thus   with    N-body
  simulations). Note that the SMT01 halo mass function best fits 
  the mass function of simulated halos that is identified with an 
  FOF linking length of 0.2 times the mean particle separation.
  The mean density of a halo so seletced is similar to that 
  within a virialized halo based on the spherical collapse model.    
  PINOCCHIO haloes and PS haloes are all defined so that 
  the mean density within a halo is similar to that based on    
  the spherical collapse model.}
\label{fig1}
\end{figure*}

\subsection{PINOCCHIO}
\label{sec:pino}

Although the problem  of  obtaining accurate  merging  histories under
ellipsoidal collapse conditions can  be circumvented by using N-body
simulations, the time-expense of these  simulations is a major hurdle. 
An  attractive  alternative  is  provided by   the LPT  code PINOCCHIO
developed recently by  Monaco \etal  (2002b).   Below we  give a short
overview of PINOCCHIO,  and we refer the  interested reader  to Monaco
\etal   (2002a,b)  and Taffoni   \etal  (2002) for   a  more elaborate
description.

PINOCCHIO uses Lagrangian perturbation theory to describe the dynamics
of  gravitational  collapse. In LPT the comoving (Eulerian) coordinate
$\bfx$ and  the initial Lagrangian coordinate $\bfq$  of each particle
are connected via
\begin{equation}
\label{displace}
\bfx(\bfq,t)= \bfq+\bfS(\bfq,t),
\end{equation}
with  $\bfS$   the  displacement   field.  The  first-order   term  of
$\bfS(\bfq,t)$ is the  well-known Zel'dovich approximation (Zel'dovich
1970):
\begin{equation}
\label{zeld}
\bfS(\bfq,t)= -D(t) {\partial \psi \over \partial \bfq}
\end{equation}
with $\psi(\bfq)$  the rescaled linear  gravitational potential, which
is related  to the  density contrast $\delta_0(\bfq)$  extrapolated to
the present time by the Poisson equation
\begin{equation}
\label{poisson}
\nabla2\psi(\bfq)= \delta_0(\bfq),
\end{equation}
Since the Lagrangian density  field is basically $\rho_{\rm L}(\bfq) =
\bar{\rho}$, the (Eulerian) density contrast is given by
\begin{equation}
\label{euldens}
1 + \delta(\bfx,t) = {1 \over {\rm det}(J)}
\end{equation}
with  $J  =  \partial  \bfx  /  \partial \bfq$  the  Jacobian  of  the
transformation  given  in~(\ref{displace}).   Note  that  the  density
formally  goes to  infinity  when the  Jacobian determinant  vanishes,
which  corresponds  to  the  point  in time  when  the  mapping  $\bfq
\rightarrow \bfx$ becomes multi-valued,  i.e.  when orbits first cross
leading  to the formation  of a  caustic.  Since  the (gravitationally
induced) flow is irrotational the matrix $J$ is symmetric and can thus
be diagonalized:
\begin{equation}
\label{euldensdiag}
1 + \delta(\bfx,t) = {1 \over \prod_{i=1}^{3}[1 - D(t) \lambda_i(\bfq)]}
\end{equation}
with   $-\lambda_i$   the  eigenvalues   of   the  deformation   tensor
${\partial2 \psi/\partial q_i \partial q_j}$.

PINOCCHIO starts  by constructing a  random realization  of a Gaussian
density field $\rho({\bfq})$ (linearly extrapolated  to $z=0$) and the
corresponding peculiar potential $\phi(\bfq)$ on   a cubic grid.   The
density fluctuation    field is  specified    completely by  the power
spectrum $P(k)$, which  is   normalized  by specifying the    value of
$\sigma_8$, defined as the rms linear  overdensity at $z=0$ in spheres
of radius $8 h^{-1} \Mpc$.  The density  and peculiar potential fields
are subsequently convolved with  a series of Gaussians with  different
values for  their FWHM $R$.  For the  $2563$ simulations used in this
paper, 26 different  linearly sampled values  of $R$ are used.  For  a
given value of $R$  the density of  a mass element (i.e.,  `particle')
will become infinite  as soon as at least  one of the ellipsoid's axes
reaches zero  size (i.e., when $D(t) =  1/\lambda_i$).  At  this point
orbit crossing (OC) occurs and the  mass element enters a high-density
multi-stream region. This is the moment of first-axis collapse.  Since
the  Jacobian determinant becomes multivalued  at  this stage, one can
not  make any  further predictions  of the  mass element's fate beyond
this point in time.  Consequently, it is  not possible in PINOCCHIO to
associate halo collapse with that of the third axis.

For each  Lagrangian point $\bfq$ (hereafter  `particle') and for each
smoothing radius $R$ this  OC (i.e., collapse)  time is  computed, and
the highest collapse redshift $z_c$, the corresponding smoothing scale
$R_c$, and  the Zel'dovich  estimate  of the  peculiar  velocity ${\bf
  v}_c$ are recorded.   PINOCCHIO differs  from  the standard  PS-like
method when it comes to assigning masses to collapsed objects.  Rather
than associating  a halo mass  with the  collapsed mass element  based
directly on the  smoothing scale $R_c$ at  collapse, PINOCCHIO  uses a
fragmentation algorithm   to link  neighboring mass   elements  into a
common dark matter halo.  In  fact, the collapsed  mass element may be
assigned to a filament or sheet rather than a halo.

After  sorting  particles according   to decreasing  collapse redshift
$z_c$ the  following rules for   accretion  and merging  are  adopted:
Whenever a particle collapses and none of its Lagrangian neighbors (the
six nearest particles) have  yet collapsed, the particle is considered
a  seed for a  new halo.  Otherwise, the  particle  is accreted by the
nearest Lagrangian neighbor that already has collapsed if the Eulerian
distance $d$, computed  using the Zel'dovich  velocities  ${\bf v}$ at
the time of collapse, obeys $d \leq f_a R_M$, where $R_M = M^{1/3}$ is
the radius  of a halo  of $M$ particles. If   more than one Lagrangian
neighbor has already  collapsed, it is simultaneously  checked whether
these haloes  merge.  This   occurs  whenever, again  at the  time  of
collapse, the mutual Eulerian distance between these haloes is $d \leq
f_M R_M$, where $R_M$ refers to the larger  halo.  Note that with this
description, up  to   six haloes may  merge  at  a given   time.   The
collapsing particles that according  to these criteria do  not accrete
onto  a halo  at their collapse  time are  assigned to a  filament. In
order to  mimic  the accretion  of   filaments onto haloes,   filament
particles can be accreted by a dark matter halo  at a later stage when
they neighbor  (in Lagrangian space) an  accreting particle.  Finally,
in high density regions  it can happen that  pairs of haloes that  are
able to merge are not touched by newly collapsing particles for a long
time.  Therefore,  at certain time  intervals pairs of touching haloes
are merged if they obey the above merging condition.

The  accretion and  merging algorithm  described above  has  five free
parameters.   In addition  to  the parameters  $f_a$  and $f_M$  three
additional  free  parameters  have  been introduced  by  Monaco  \etal
(2002b).   We  refer the  reader  to  this  paper for  details.   This
relatively large amount  of freedom may seem a  weakness of PINOCCHIO. 
However, it is  important to realize that even  N-body codes require
some   free   parameters,   such   as  the   linking-length   in   the
Friends-Of-Friends  (FOF)  algorithm  used  to  identify  dark  matter
haloes.  Furthermore, we  do not consider these parameters  as free in
what follows.  Rather,  we adopt the values advocated  by Monaco \etal
(2002a,b), 
which  they obtained by  tuning PINOCCHIO to  reproduce the
conditional and unconditional mass function of N-body simulations.

\begin{figure}
\centerline{\psfig{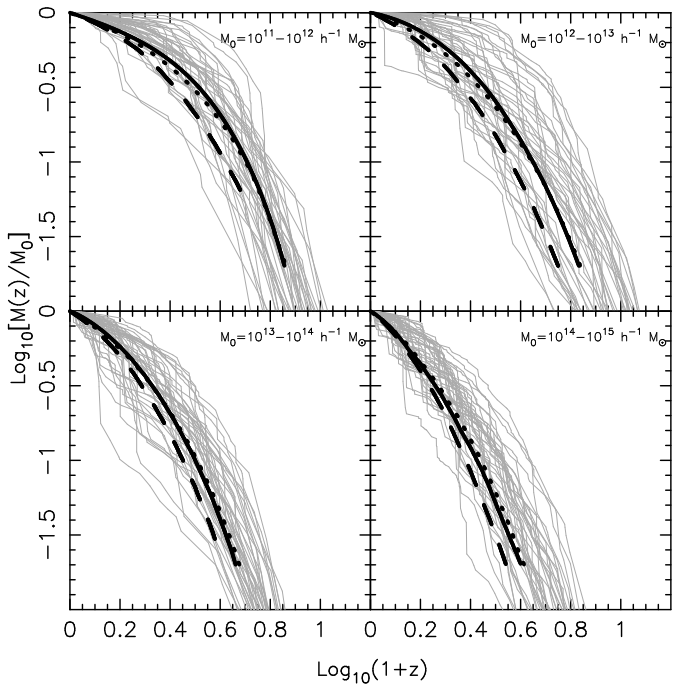}}
\caption{The mass assembly histories of dark matter haloes
  with present-day masses in the four mass bins as indicated 
  in the panels. The upper two panels are based on the 
  $100 h^{-1}{\rm Mpc}$-box simulations, P1 and S1, 
  while the lower two panels use data from the 
  $300 h^{-1}{\rm Mpc}$-box simulations, P2 and S2.
  The thin lines are 40 MAHs randomly selected from the PINOCCHIO 
  simulations. The thick solid line in each panel shows the  average 
  of all the  MAHs obtained in the PINOCCHIO simulaions in the 
  corresponding mass bin. The thick dotted line shows the average
  MAH extracted from the simulations. The thick dashed line
  shows the average MAH obtained from 3000 EPS realizations 
  (properly  sampled from halo mass function).}
\label{MAH}
\end{figure}

\begin{figure}
\centerline{\psfig{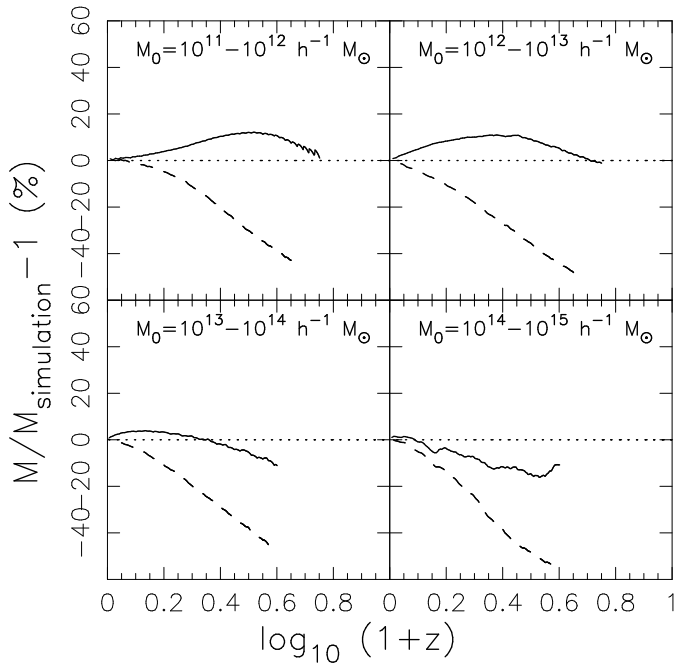}}
\caption{The dashed curve in each panel shows the difference 
  between  the average MAHs predicted  by  the  EPS  model and by 
  the N-body simulation,
  while the solid    curve shows the difference  between
  PINOCCHIO prediction and N-body simulation. 
  The the upper two panels use data from P1 and S1, while the lower two 
  panels use data from P2 and S2. 
  Data are not shown for $z \gsim 3$  because the MAHs are  
  not well represented at  such high redshifts in the simulations.}
\label{MAH2}
\end{figure}

{\bf
\begin{figure}
\centerline{\psfig{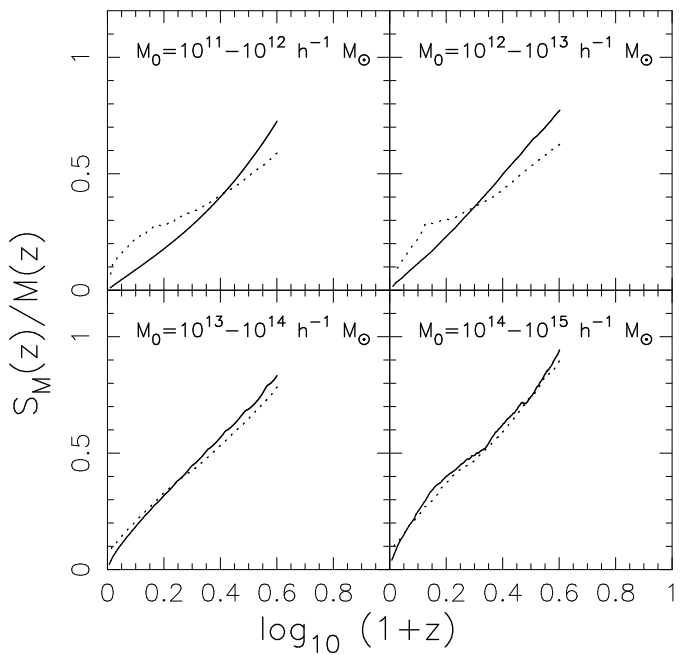}}
\caption{The {\it standard deviation} of the MAHs, $S_{\rm M}(z)$, 
normalized by the average MAH, $M(z)$, in four mass bins. 
Solid lines are results from PINOCCHIO, while dotted lines 
are results from N-body simulations.
As in Fig.~\ref{MAH} and Fig.~\ref{MAH2}, the upper two panels
use data from P1 and S1, while the lower two panels use data
from P2 and S2.}
\label{scatter}
\end{figure}
}
\begin{table}
\begin{center}
\caption{Ensemble of PINOCCHIO simulations (P0)}
\begin{tabular}{cccc}
\hline
Box size ($h^{-1}$ Mpc)& $N_{\rm run}$ & $M_{p}$ ($h^{-1}\Msun$) & $N_{\rm MAH}$ \\
\hline
 20 & 12 & $4.0 \times 107$    & 2,690 \\
 40 &  8 & $3.2 \times 108$    & 1,863 \\
 60 &  8 & $1.1 \times 109$    &   796 \\
 80 &  6 & $2.5 \times 109$    & 1,438 \\
100 &  6 & $5.0 \times 109$    & 2,799 \\
140 &  4 & $1.4 \times 10^{10}$ &   410 \\
160 &  2 & $2.0 \times 10^{10}$ &   299\\
200 &  9 & $4.0 \times 10^{10}$ & 2,629 \\ 
\hline
\end{tabular}
\end{center}
\medskip
A  listing of  the  PINOCCHIO  simulations used  in  this paper.   All
simulations use $2563$ particles  and adopt the standard $\Lambda$CDM
concordance  cosmology. In order to get good statistics, we choose a
combination of box sizes so that we can select thousands of 
well-resolved (with more than 2000 particles) haloes 
in each mass bin we adopt in the paper. This
ensemble of PINOCCHIO simulations is referred to as `P0' in the text.
The  first column  of Table 1 lists the  box size  of the
simulation in  $h^{-1} \Mpc$.  The  second column lists the  number of
independent  realizations run.   The particle  mass $M_p$  (in $h^{-1}
\Msun$) is listed  in the third column, while  the fourth column lists
the   total  number  of   haloes  (summed   over  all   $N_{\rm  run}$
realizations) with  more than 2000 particles  and for which  a MAH has
been obtained.
\end{table}

\section{Simulations}
\label{sec:sim}

In this paper we use PINOCCHIO simulations to  study the mass assembly
histories (MAHs) of dark  matter  haloes.  We follow  previous studies
(Lacey \& Cole  1993; Eisenstein \& Loeb  1996; Nusser \& Sheth  1999;
van den Bosch 2002a) and define the MAH, $M(z)$, of a halo as the main
trunk of  its   merger tree: at    each redshift, the   mass $M(z)$ is
associated   with  the mass  of the  most   massive progenitor at this
redshift,  and  we follow this progenitor,   and this progenitor only,
further back in time.  In this way,  this `main progenitor halo' never
accretes other  haloes that are more massive  than itself.   Note that
although  at each branching point we   follow the most massive branch,
this does not necessarily imply  that the main  progenitor is also the
most massive of {\it all} progenitors at any given redshift.

Below we describe the PINOCCHIO simulations, the N-body simulations,
and the EPS method used  to construct MAHs. 

\subsection{PINOCCHIO simulations}
\label{sec:pinsim}

Because the progenitors of a present-day halo become smaller at higher
redshift, we can only follow the  MAHs to a sufficiently high redshift
if the  halo  at $z=0$ contains  a  large enough number of  particles. 
When constructing MAHs with PINOCCHIO, we only use haloes that contain
more than 2000 particles at the present time, and we trace each MAH to
the redshift  at  which  its main  progenitor  contains  less than  10
particles.  In order to cover  a large range  of halo masses, we have
carried out 55  PINOCCHIO simulations with  $2563$ particles each and
spanning a wide range of box sizes  and particle masses (see Table~1, we 
call this suite of PINOCCHIO simulations P0 hereafter). 
The choice of box sizes ensures that there are several thousand
well-resolved haloes in each of the mass bins considered.
Each  of these simulations  takes only about 6  hours of CPU time on a
common PC (including the actual  analysis), clearly demonstrating  its
advantage over regular N-body simulations. This suite of PINOCCHIO
simulations has adopted
the $\Lambda$CDM    concordance   cosmology   with   $\Omega_m=0.3$,
$\Omega_\Lambda=0.7$, $h=0.7$ and $\sigma_8=0.9$.

With simulation box  sizes ranging from $20 \mpch$  to $200\mpch$, and
particle  masses ranging  from $4  \times 10^{7}  h^{-1} \msun$  to $4
\times  10^{10}  h^{-1} \msun$,  we  are able  to  study  the MAHs  of
present-day haloes with masses $> 8 \times 10^{10} h^{-1} \msun$.  The
construction of the MAHs  is straightforward: PINOCCHIO outputs a halo
mass every time  a merger occurs, i.e., when a halo  with more than 10
particles merges into  the main branch.  If we  require an estimate of
the  halo  mass at  any  intermediate  redshift,  $z$, we  use  linear
interpolation   in  $\log(1+z)$  between   the  two   adjacent  output
redshifts.
  
\subsection{N-body simulations}
\label{sec:nbody}

  For comparison we also used MAHs extracted from two sets of N-body
simulations (referred to as S1 and S2). These N-body simulations
follow the evolution of $5123$  particles in a periodic box of  
$100 \mpch$ (S1) and $300 \mpch$ (S2) on a side, assuming slightly 
different cosmologies (see Table 2 for details).
The snapshot outputs of each simulation are evenly placed at 60 
redshifts between $z=0$ and $z=15$ in $\ln(1+z)$ space. 

In each simulation and at each output, haloes are identified using the
standard  FOF algorithm  with  a linking  length  of $b=0.2$.   Haloes
obtained with  this linking  length have a  mean overdensity  of $\sim
180$.  A  halo at redshift  $z_1$ is identified  as a progenitor  of a
halo at $z_2 <  z_1$ if more than half of  its mass is included in
the halo at $z_2$. 
The resulting  lists of  progenitor haloes are  used to  construct the
MAHs.   In  our  analysis,  we  only  use  haloes  more  massive  than
$10^{11}h^{-1}\Msun$ at the present time in S1 and 
halos more massive than $10^{13}h^{-1}\Msun$ in S2. Thus,
in each simulation only halos with more than $\sim 600$ particles 
at $z=0$ are used, which allows us to trace the  MAHs to sufficiently high
redshift with sufficiently high resolution. For comparison, we also
generate two sets of PINOCCHIO simulations, P1 and P2, 
using exactly the same numbers of particles and cosmologies as 
in S1 and S2, respectively (see Table 2).

\subsection{Monte-Carlo simulations}
\label{sec:moncar}

We  also generate  MAHs  using Monte-Carlo  simulations  based on  the
standard  EPS  formalism.  We  adopt  the  N-branch  tree method  with
accretion suggested  by Somerville \&  Kolatt (1999, hereafter  SK99). 
This method yields more reliable MAHs than for example the binary-tree
method of Lacey \& Cole  (1993).  In particular, it ensures exact mass
conservation, and  yields conditional mass functions that  are in good
agreement with direct predictions from EPS theory (i.e., the method is
self-consistent).

To construct  a merger  tree for a  parent halo  of mass $M$  the SK99
method works as  follows.  First a value for $\Delta  S$ is drawn from
the mass-weighted probability function
\begin{equation}
\label{probdS}
P(\dS ,\dW)  \; {\dd}\dS = {1 \over \sqrt{2  \pi}} \; {\dW \over
\dS^{3/2}} \; {\rm exp}\left[-{(\dW2) \over 2 \dS}\right] \; {\dd}\dS
\end{equation}
(cf. equation~[\ref{probSS}]).  Here  $\dW$ is a  measure for the time
step  used in the merger  tree, and is a   free parameter (see below). 
The progenitor mass,  $M_p$,    corresponding to $\dS$ follows    from
$\sigma2(M_p) = \sigma2(M) + \dS$.   With each new progenitor it  is
checked   whether the sum  of   the progenitor  masses drawn thus  far
exceeds   the  mass of the  parent,  $M$.   If  this  is  the case the
progenitor    is rejected and a  new   progenitor mass  is drawn.  Any
progenitor  with $M_p  < M_{\rm min}$  is  added to the mass component
$M_{\rm acc}$ that is  considered to be accreted  onto the parent in a
smooth fashion  (i.e.,   the formation  history of  these   small mass
progenitors is not followed further back in time).  Here $M_{\rm min}$
is a free  parameter that has to be  chosen sufficiently small.   This
procedure is repeated until  the total mass left, $M_{\rm  left} = M -
M_{\rm acc}  - \sum M_p$, is less  than $M_{\rm min}$.  This remaining
mass is assigned to $M_{\rm  acc}$ and one  moves on to the next  time
step. For  the construction of MAHs, however,  it  is not necessary to
construct an  entire set of  progenitors.  Rather, at each  time step,
one can stop once the  most massive progenitor  drawn thus far is more
massive than $M_{\rm  left}$.  This has  the additional advantage that
one does not  have to define a minimum  progenitor mass  $M_{\rm min}$
(see van den Bosch 2002a for details).
 
In principle, since the  upcrossing of trajectories through a boundary
is a  Markov process,  the statistics of  progenitor masses  should be
independent of the  time steps taken.  However, the  SK99 algorithm is
based on the  {\it single} halo probability (equation~[\ref{probdS}]),
which  does  not  contain  any  information about  the  {\it  set}  of
progenitors that make up the  mass of $M$.  In fact, mass conservation
is enforced  `by hand', by  rejecting progenitor masses  that overflow
the mass budget.  As shown in van den Bosch (2002a), this results in a
time step  dependency, but only  for relatively large time  steps. For
sufficiently small values of $\dW$ the algorithm outlined above yields
accurate and robust results (see also SK99).  Throughout this paper we
adopt a timestep  of $\Delta z=0.05$.  Our tests with different 
values of $\Delta z$ from $0.01$ to $0.05$  have shown that this
time step is small enough to achieve stable results, that is, when we
decrease the time step to $\Delta z=0.01$, the change in the 
average MAH is less than 1\%.

\subsection{Comparison}
\label{sec:comp}

\begin{table*}
\begin{center}
\caption{Reference PINOCCHIO and N-body simulations}
\begin{tabular}{lccccccc}
\hline\hline
Simulation Name & $N_{\rm p}$ &Box size  ($h^{-1}$ Mpc) & $M_{p} (h^{-1}\Msun)$ &
$\Omega_{\rm m}$ & $\Omega_\Lambda$ & $h$ & $\sigma_8$ \\
\hline\hline
S1 (N-body) & $5123$ &100 & $5.5 \times 10^{8}$ &  0.268 & 0.732 & 0.71 & 0.85\\
P1 (PINOCCHIO)& $5123$ &100 & $5.5 \times 10^{8}$ & 0.268 & 0.732 & 0.71 & 0.85\\
\hline
S2 (N-body)& $5123$ &300 & $1.3 \times 10^{11}$ & 0.236 & 0.764 & 0.73 & 0.74 \\
P2 (PINOCCHIO)& $5123$ &300 & $1.3 \times 10^{11}$ & 0.236 & 0.764 & 0.73 & 0.74\\
\hline\hline
\end{tabular}
\end{center}
\medskip
\end{table*}
We now compare   the MAHs obtained  with all  three methods  discussed
above.   The upper panels  of Fig.~\ref{fig1} plot the (unconditional)
halo  mass  functions at    four  different redshifts,   as indicated,
obtained from 5 arbitrary   PINOCCHIO runs with different box  sizes in P0. 
Dashed lines correspond to the analytical halo mass functions obtained
using the standard PS formalism (equation~[\ref{PS}]), while the solid
lines   indicate  the mass functions  of  SMT01   based on ellipsoidal
collapse.  The  latter have  been  shown to accurately  match the mass
functions  obtained from N-body simulations (e.g., Sheth \& Tormen, 1999;
SMT01). The symbols in  the
lower  panels  of  Fig.~\ref{fig1}  plot the   differences between the
PINOCCHIO and  the   SMT01 mass  functions,  while  the  dashed  lines
indicate the differences between the PS and  the SMT01 mass functions. 
Clearly, the PINOCCHIO mass  functions are in excellent agreement with
those  of  SMT01, and  thus  also  with  those  obtained from N-body
simulations.  In  addition,  Taffoni  \etal  (2002)  have  shown  that
PINOCCHIO also accurately matches the {\it conditional} mass functions
obtained from  numerical simulations.  We  now investigate whether the
actual MAHs obtained  from PINOCCHIO are also  in good  agreement with
the numerical simulations.

 Fig.~\ref{MAH} plots the average MAHs obtained from the PINOCCHIO, 
N-body and EPS simulations, for halos with the present masses 
in the following four mass ranges: 
$\log(M_0/h^{-1}\msun)=$11-12, 12-13,  13-14 and  14-15.
For comparison, in each panel we also show 40 randomly selected  
MAHs from the PINOCCHIO simulations (P1 and P2). 
To ensure mass resolution, results for the low-mass bins
(the two upper panels) are based on simulations with the small 
box size, i.e. S1 and P1. Results for the high-mass bins
(the two lower panels) are based only on simulations with the 
large-box size (S2 and P2) in order to obtain a large number of 
massive halos. The  thick solid  curve in each panel corresponds  
to the average MAH obtained by averaging over all the halos,
in the mass range indicated, found in one of the PINOCCHIO simulations 
(P1 and P2). The thick dashed lines correspond to the average MAHs 
obtained from 3000 EPS Monte-Carlo simulations (properly weighted by 
the halo mass function). The thick dotted lines show the average MAHs 
obtained from the two N-body simulations (S1 and S2).
In Fig.~\ref{MAH2}, a detailed comparison between these 
results are  presented. As can be seen in 
Fig.~\ref{MAH2}, the average MAHs obtained with  PINOCCHIO are in good
agreement with  those  obtained from  the N-body simulations  (with
differences  smaller than 10\%). Note that there are uncertainties
in the identification of dark haloes in N-body simulations
using the FOF algorithm. Sometimes two physically separated haloes 
can be linked together and identified as one halo if they 
are bridged by dark matter particles, which can change 
the halo mass by 5\% on average. The agreement between 
PINOCCHIO and simulation shown in Fig.~\ref{MAH2} is 
probably as good as one can hope for. 
The EPS model, however,   yields  MAHs that are systematically
offset  with respect to those  obtained from the N-body simulations:
the  EPS  formalism  predicts that  haloes  assemble   too  late
(see also van den Bosch 2002a; Lin,  Jing \& Lin 2003; W02). 
Fig.~\ref{scatter} shows the ratio between the standard deviation 
of the MAHs, $S_{\rm M}(z)$, and the average MAH $M(z)$, as a function 
of redshift $z$. As one can see, the agreement between the PINOCCHIO 
and N-body simulations is also reasonably good.

In  summary, the  Lagrangian Perturbation  code PINOCCHIO  yields halo
mass  functions (both conditional  and unconditional), and 
 mass assembly histories that  are  all  in  good
agreement  with N-body  simulations.  In  particular, it  works much
better than the  standard PS formalism, and yet is  much faster to run
than numerical simulations. PINOCCHIO  therefore provides a unique and
fast platform  for accurate investigations of  the assembly histories
of a large, statistical sample of CDM haloes.

\section{Halo formation times}
\label{sec:ftime}

Having demonstrated that the PINOCCHIO MAHs are in good agreement with
those obtained from N-body simulations, we now  use the suite of 55 PINOCCHIO
simulations, P0,  listed in Table~1 to investigate the assembly histories of
a large sample of haloes spanning a wide range in halo masses.

The assembly history  of a  halo can be  parameterized by  a formation
time (or  equivalently  formation redshift), which  characterizes when
the  halo assembles.   However,  since the  assembly  of  a halo is  a
continuous  process, different `formation  times' can be defined, each
focusing on a different aspect of the MAH.  Here we define and compare
the following four formation redshifts:

\begin{enumerate}
  
\item  $z_{\rm half}$:  This is  the redshift  at which  the  halo has
  assembled  half of  its final  mass.  This formation  time has  been
  widely used in the literature.
  
\item $\zlmm$: This is redshift at which the halo experiences its last
  major  merger.  Unless stated otherwise we  define a major merger as
  one in  which the mass ratio  between the two progenitors  is larger
  than $1/3$. This definition is  similar to $z_{\rm jump}$ defined in
  Cohn \& White (2005). Major mergers may have played an important 
  role in transforming galaxies and in regulating star formation in 
  galaxies. Their frequency is therefore important to quantify. 
  
\item $\zvczero$: This is the redshift at which the virial velocity of
  a  halo, $\Vh$,  defined  as  the circular  velocity  at the  virial
  radius, reaches its current value, $V_0$, for the first time.  Since
  $\Vh$ is a  measure for the depth of  the potential well, $\zvczero$
  characterizes  the  formation   time  of  the  halo's  gravitational
  potential.
  
\item  $\zvcpeak$: This  is the  redshift at  which the  halo's virial
  velocity reaches its maximum value  over the entire MAH.  As we show
  below, the  value of $\Vh$  is expected to increase  (decrease) with
  time, if the time scale  for mass accretion is shorter (longer) than
  the  time  scale of  the  Hubble  expansion.  Therefore,  $\zvcpeak$
  indicates the time when the MAH transits from a fast accretion phase
  to a slow accretion phase.

\end{enumerate}

In an N-body simulation one can infer the virial  velocity of a halo
from  its  internal structure.  In the case  of PINOCCHIO simulations,
however, no information regarding  the density distribution  of haloes
is available. However, we may use the fact that CDM haloes always have
a particular  (redshift and   cosmology dependent) overdensity.   This
allows us to define the virial velocity at redshift $z$ as
\begin{equation}
\label{eq:vcz}
\Vh(z) = \sqrt{G \Mh \over \Rh} =
\left[ \frac{\deltac(z)}{2}\right]^{1/6} \left[\Mh(z) \, H(z)\right]^{1/3}
\end{equation}
Here $\Mh$  and $\Rh$  are the  virial mass and  virial radius  of the
halo, respectively, and $H(z)$  is the Hubble parameter.  The quantity
$\deltac(z)$ is the  density contrast between the mean  density of the
halo  and the  critical  density for  closure,  for which  we use  the
fitting formula of Bryan \& Norman (1998),
\begin{equation}
\label{delc}
\deltac(z) = 18 \pi2 + 82 [\Omega_{\rm m}(z)-1] - 39 [\Omega_{\rm m}(z)-1]2 
\end{equation}

\begin{figure}
\vbox{
\psfig{file=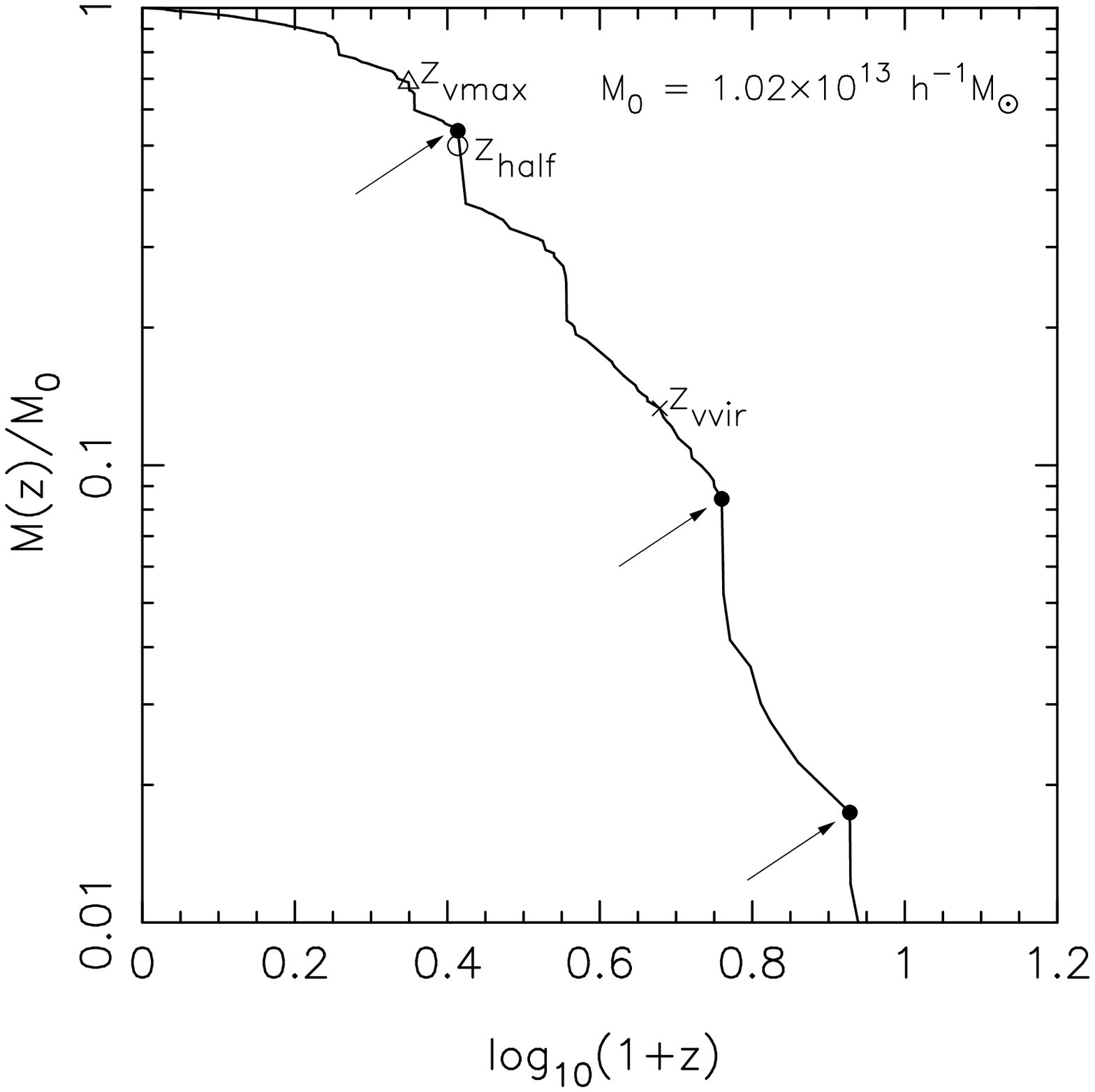,angle=270,width=1.0\hsize} 
\psfig{file=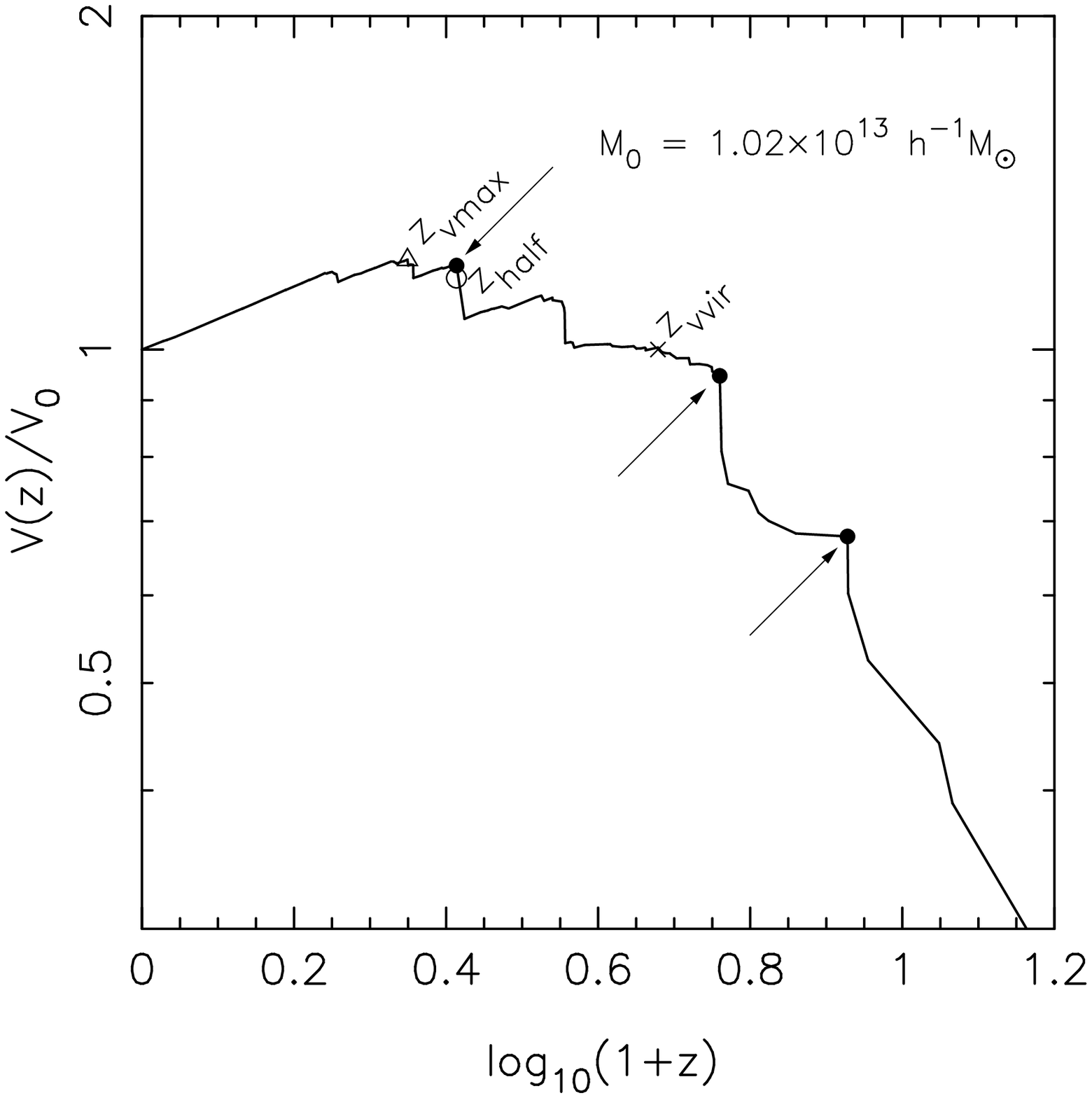,angle=270,width=1.0\hsize}
\caption{{\it Upper panel:} the MAH of a randomly chosen halo with a 
  mass of  $1.02 \times 10^{13}  h^{-1}\Msun$.  Various characteristic
  events during the assembly   of this halo are  indicated: $\zvcpeak$
  (open triangle), $\zhalf$ (open circle), and $\zvczero$ (cross). The
  solid dots with  an arrow indicate  major mergers (those with a mass
  ratio larger than $1/3$). {\it Lower panel:} same as in upper panel,
  except   that here the  evolution  of  the halo   virial velocity is
  shown.}
\label{fig:zformex}
}
\end{figure}

As  an illustration, Fig.~\ref{fig:zformex}  plots the MAH, $M(z)/M_0$
(upper  panel), and the history of   the virial velocity, $\Vh(z)/V_0$
(lower  panel) for a randomly  selected halo (with  $M_0 = 1.02 \times
10^{13} h^{-1} \Msun$).  All major merger events are marked by a solid
dot plus arrow.  The  last major merger occurs at  $\zlmm= 1.60$.  The
other  formation  redshifts,  $\zhalf=1.59$,   $\zvczero=3.77$,    and
$\zvcpeak=1.23$ are marked  by an open  circle,  a cross, and  an open
triangle, respectively.

\begin{figure}
\centerline{\psfig{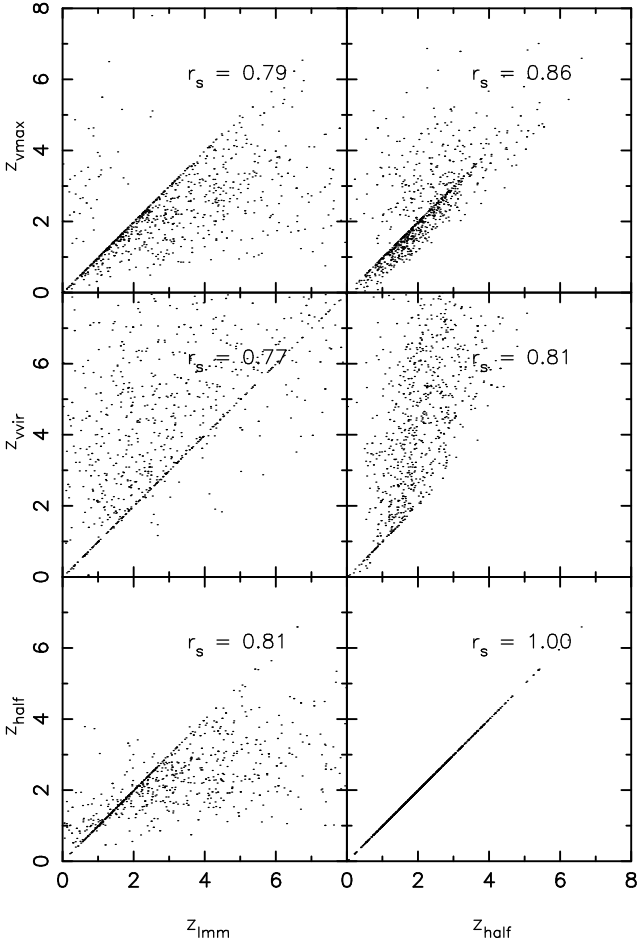}}
\caption{The correlations between various halo formation redshifts for 
  haloes with  present day  masses in  the range  $10^{11} h^{-1} \Msun
  \leq M \leq 10^{12} h^{-1} \Msun$.  The value of $r_s$ in each panel
  shows the corresponding Spearman rank-order correlation coefficient.
  Due to the finite  time resolution in  the PINOCCHIO simulations, in
  some cases the values of two formation times can be the same.}
\label{fig:zformcorr}
\end{figure}

Fig.~\ref{fig:zformcorr}  plots the  correlations between  the various
formation redshifts, for  haloes with masses  in the range $10^{11}  -
10^{12} h^{-1}\Msun$.   The value  of $r_s$  in each  panel  shows the
corresponding Spearman  rank-order correlation coefficients.  Clearly,
there is  significant  correlation among all the  formation redshifts,
but the scatter is quite large. This demonstrates that these different
formation times characterize different  aspects of a  given MAH.  
Unlike simulation which outputs snapshots at arbitrary times, 
PINOCCHIO only outputs when a merger occurs and the merger is treated as 
instantaneous. Consequently, some formation times can have exactly 
the same value in PINOCCHIO simulations. Note
that the correlation shown in the lower left panel is quite similar to
that obtained  by Cohn   \& White  (2005)  for simulated  clusters  of
galaxies.   Note also that typically, $\zvczero  > \zhalf$ and $\zvczero
> >\zlmm$.  This shows that haloes {\it  in this mass range} established
their potential wells  before they accreted  a major fraction of their
mass.  The last major merger typically  occurred well before $\zhalf$,
which indicates that most  of that mass has been  accreted in a fairly
smooth fashion (see also W02 and Zhao \etal 2003a).

\begin{figure}
  \centerline{\psfig{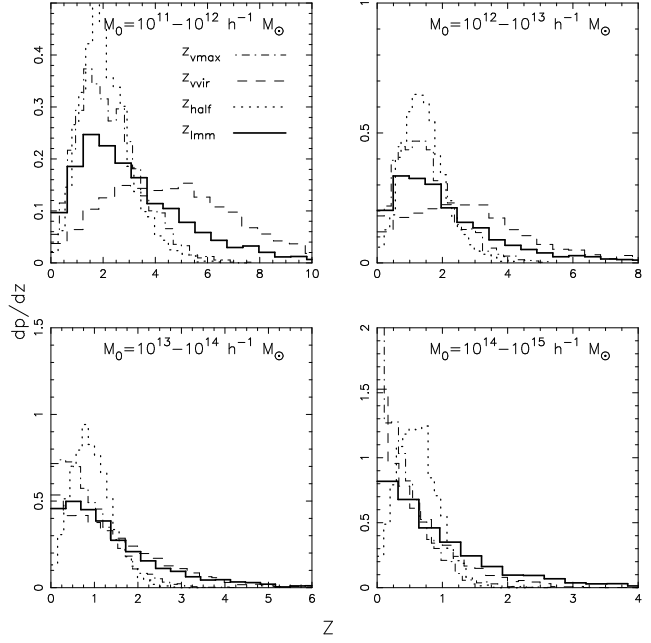}}
\caption{The probability distributions of $\zhalf$ (dotted lines), 
  $\zvczero$ (dashed lines), $\zvcpeak$ (dot-dashed lines) and $\zlmm$
  (thick solid lines). Results are shown for four different mass bins,
  as indicated in  each panel. Note that  the scale of the four panels
  is different!  See text for a detailed discussion.}
\label{fig:zform1}
\end{figure}

\begin{figure}
\centerline{\psfig{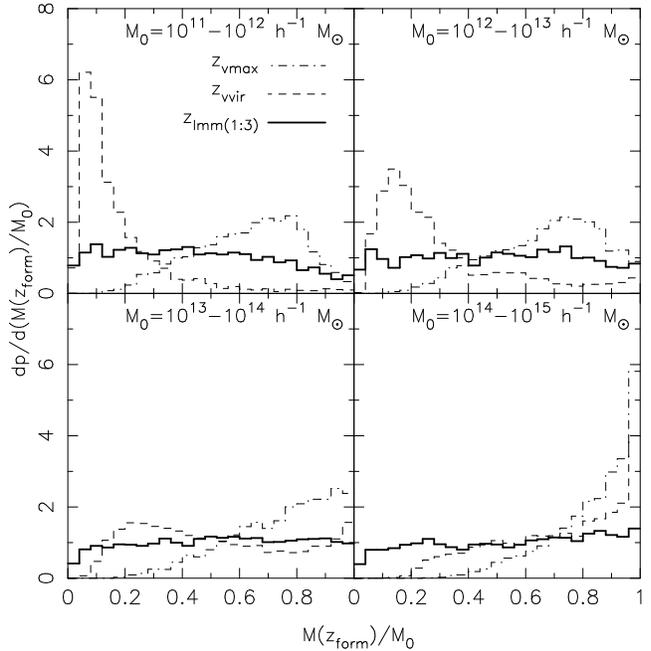}}
\caption{The distributions of the halo mass fraction at various formation 
  times. Different line-styles correspond  to different definitions of
  the formation time, as indicated in the upper left-hand panel. As in
  Fig.~\ref{fig:zform1}, different panels correspond to different halo
  mass bins, as indicated.}
\label{fig:zform2}
\end{figure}

Fig.~\ref{fig:zform1}  shows the distributions  of  the four formation
redshifts defined above.  Results   are shown for four  different mass
bins, as  indicated. For all four  formation redshifts, the  median is
higher for haloes  of  lower masses.   This reflects  the hierarchical
nature of the  assembly of dark matter   haloes: less massive  systems
assemble  (`form') earlier.  Note that   the distribution of formation
times is also  broader for  lower mass haloes.   For haloes  with $M_0
\gta M^{*} \simeq 10^{13}  h^{-1} \Msun$\footnote{Here $M^{*}$  is the
  characteristic  non-linear   mass   defined  by   $\sigma(M^{*})   =
  \delta_{\rm crit}0$}, all the distribution functions except that of
$z_{\rm half}$ are peaked  at, or very near  to, $z = 0$.   This shows
that the majority of  these haloes are still  in their  fast accretion
phase,  so that their potential wells  are still  deepening with time. 
On  the  other  hand,  haloes  with $M_0  \ll   M^{*}$ typically  have
$\zvczero    >       \zhalf$    and      $\zvczero     >\zlmm$    (cf. 
Fig.~\ref{fig:zformcorr}), indicating  that their potential wells have
already been  established, despite the  fact   that they  continue  to
accrete appreciable amounts of mass.

Fig.~\ref{fig:zform2} shows  the distributions of  the ratio $M(z_{\rm
  form})  /  M_0$, with  $z_{\rm  form}$  one  of our  four  formation
redshifts.  By definition, the distribution  of $M(\zhalf) / M_0$ is a
$\delta$-function at $M(z_{\rm form})/M_0 = 0.5$, and is therefore not
shown.   For haloes  with $M_0  <  10^{13} h^{-1}  \Msun$, the  virial
velocity has already  reached the present day value  when the halo has
only  assembled 10\%-20\%  of  its final  mass.   Thus, these  systems
assemble most of  their mass without significant changes  to the depth
of  their potential  well.  Only  for  massive haloes  with $M_0  \gta
10^{14} h^{-1} \Msun$ is the median of $M(\zvczero) / M_0$ larger than
0.5, implying that  they have assembled the majority  of their present
day mass through major (violent) mergers.

If we define major mergers as those with a  progenitor mass ratio that
is at least $1/3$, the  distribution  of $M(\zlmm)/M_0$ is  remarkably
flat. This implies  that some  haloes  accrete a large  amount of mass
after their last major merger, while  for others the last major merger
signals  the last significant mass  accretion  event.  Remarkably, the
distribution of $M(\zlmm)/M_0$ is virtually independent of $M_0$.  For
low mass  haloes, the flatness  of  the distribution of $M(\zlmm)/M_0$
simply reflects  the   broad distribution  of  $\zlmm$.  However,  for
massive haloes  with $M  \gta M^{*}$,  the distribution of  $\zlmm$ is
fairly narrow.   Therefore,  for  these  haloes the  flatness  of  the
$M(\zlmm)/M_0$ distribution   implies  that, since   their last  major
merger,  they have accreted a significant  amount of mass due to minor
mergers. Since the last major merger occurred fairly recently, this is
another  indication that massive   haloes   are still in their    fast
accretion phase.

\section{The properties of major mergers}
\label{sec:majmerprop}

During  the assembly  of dark  matter  haloes, major  mergers play  an
important role. Not only does  a major merger add a significant amount
of mass,  it also deepens  the halo's potential well.  Furthermore, in
current models of galaxy formation, a major merger of two galaxy-sized
haloes  is  also expected  to  result in  a  merger  of their  central
galaxies, probably triggering a starburst and leading to the formation
of an  elliptical galaxy. Therefore,  it is important to  quantify the
frequency of major mergers during the formation of CDM haloes.

\begin{figure}
\centerline{\psfig{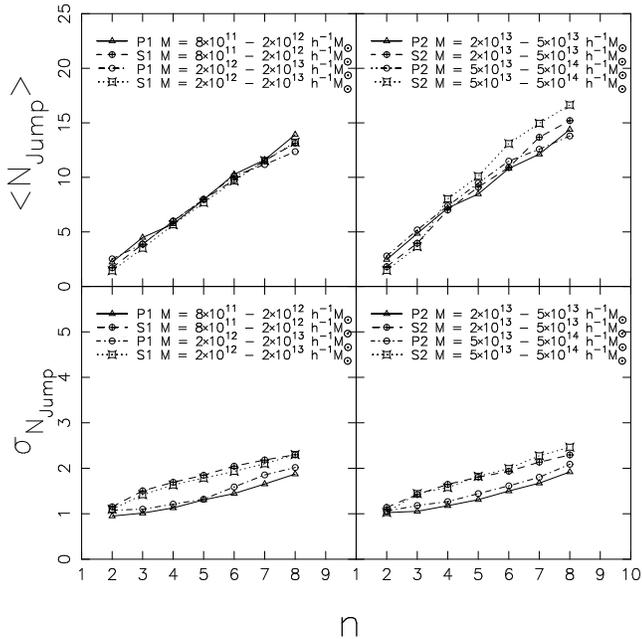}}
\caption{The median, $\langle N_{\rm jump}\rangle$, and 
dispersion, $\sigma_{N_{\rm jump}}$, of the distribution 
of the number of mass jumps, $N_{\rm jump}$, in the MAHs, 
versus $n$ (see text for definitions). Left  panels show
comparison between P1 and S1, while right panels 
show comparison between P2 and S2. Note that the agreement 
between the PINOCCHIO simulations and $N$-body simulations 
is remarkable and the mass dependence is rather weak.}
\label{njump}
\end{figure}

 As mentioned above, in a PINOCCHIO simulation mergers of dark matter
haloes are treated as instantaneous events, and the masses of the 
merger progenitors are recorded whenever a merger happens.
This makes it very convenient to identify mergers in PINOCCHIO.
On the other hand, in an $N$-body simulation halos are identified only 
in a number of snapshots, and so the accuracy of identifying mergers is 
limited by the time intervals of the snapshots. For example, 
if we define major mergers by looking for halos for which   
the mass ratio between its second largest and largest progenitors 
exceeds 1/3 in the last snapshot, we may miss major mergers in which 
the two progenitors were assembled during the two snapshots.
On the other hand, if we identify major mergers in a simulation 
by looking for halos whose masses increase by a factor between 
1/4 and 1 in the next snapshot, we will overestimate 
the number of major merger events, because some of the halos 
may have increased their masses by accretion of small halos
rather than through major mergers. In the simulations 
used here (S1 and S2), the time intervals between successive 
snapshots are about 0.3-0.6 Gyr, comparable to the time scales of 
major mergers, and the two definitions of major mergers described 
above lead to a factor of 2 difference in the number of 
major mergers. Because of this, it is difficult to make a direct 
comparison between PINOCCHIO and N-body simulations in their 
predictions for the number of major mergers. In order to check 
the reliability of PINOCCHIO in predicting the number of 
major mergers, we use quantities that are related to the 
number of major mergers but yet can be obtained from both 
our N-body and PINOCCHIO simulations. We first construct PINOCCHIO 
haloes at each of the snapshots of our N-body simulations.  
We then follow the MAH of each of the present halo using 
the snapshots and identify the number of events in which 
the mass of a halo increases by a factor exceeding $1/n$ 
between two successive snapshots, where $n$ is an integer 
used to specify the heights of the jumps. In practice, 
we trace the MAH backward in time until the mass of the halo 
is 1\% of the final halo mass. Since exactly the same analysis 
can also be carried out for the N-body simulations, we can 
compare, for a given $n$ and for halos of given mass at the present 
time,  the statistics of the number of jumps, $N_{\rm jump}$,
predicted by PINOCCHIO simulations with that given by
the N-body simulations. We found that the distribution 
of $N_{\rm jump}$ for a given $n$ can be well fit by a Gaussion 
distribution, and in Fig.~\ref{njump} we plot the median 
$\langle N_{\rm jump}\rangle $ and standard deviation 
$\sigma_{N_{\rm jump}}$ versus $n$, in several mass bins. 
The agreement between PINOCCHIO and N-body simulations is 
remarkably good. Although $N_{\rm jump}$ is not exactly 
the number of major mergers, the good agreement between 
PINOCCHIO and N-body simulations makes us believe that 
it is reliable to use PINOCCHIO to make predictions for the 
statistics of major mergers.  

\begin{figure}
\centerline{\psfig{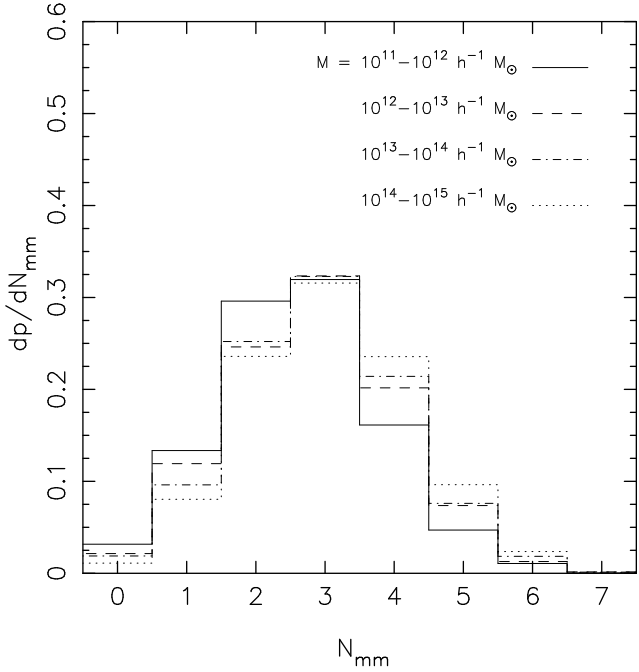}}
\caption{The distribution of the number of major mergers (those with a  
  mass  ratio  larger  than  $1/3$)  in   our PINOCCHIO   simulations. 
  Lines in different styles represent different mass bins.
  Note that the distributions are virtually independent of
  halo mass.}
\label{mm}
\end{figure}

\begin{figure}
\centerline{\psfig{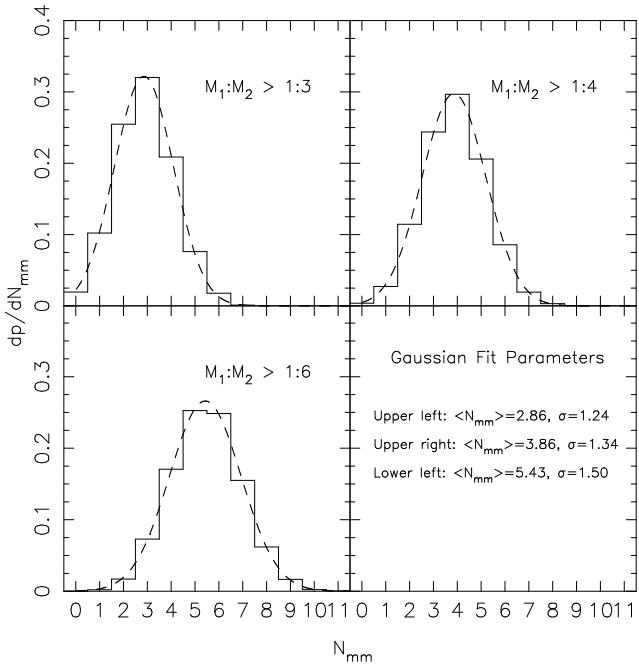}}
\caption{Distribution of the number of mergers (in PINOCCHIO simulations) 
  with a mass ratio  larger than $1/3$  (upper left-hand panel), $1/4$
  (upper right-hand panel), and $1/6$ (lower left-hand panel).  In all
  three cases all haloes with masses in the range from $10^{11} h^{-1}
  \Msun$ to $10^{15} h^{-1}\Msun$ are used. The dotted curves show the
  best-fit Gaussians,  the median and standard  deviation of which are
  indicated in the lower right-hand panel.}
\label{fig:mmstat}
\end{figure}

 In order to investigate the statistics of major mergers in detail, 
we count the number of major mergers for each of the halos
in the ensemble of simulations P0. Here again we only trace 
a halo back to a time when the mass of its main progenitor
is  1\% of the halo's final mass. This  choice of lower mass limit 
is  quite arbitrary. However, some limit is necessary, because otherwise 
there  will be a large number of major mergers involving  progenitors 
with excessively small masses  at very early times. 
Furthermore this mass limit is also the one we use in defining 
$N_{\rm jump}$. The large number of halos in the ensemble ensures 
that each mass bin contains about 2000 haloes. 
Fig.~\ref{mm} plots the distributions of the number
of major  mergers (with a progenitor  mass ratio $\ge 1/3$) for haloes
of different masses  at the present time.  A  halo experiences about 1
to 5 major  mergers during its mass assembly  history, with an average
of about  3.  Note that  the $N_{\rm mm}$-distributions  are virtually
independent       of   halo   mass.      As    we    have  shown    in
Section~\ref{sec:ftime}, however, the redshifts at which these mergers
occur do depend strongly on halo  mass: while most major mergers occur
before  $z  \simeq 2$ for galaxy-sized  haloes,  they occur  much more
recently in the more massive, cluster-sized haloes.

\begin{figure}
\centerline{\psfig{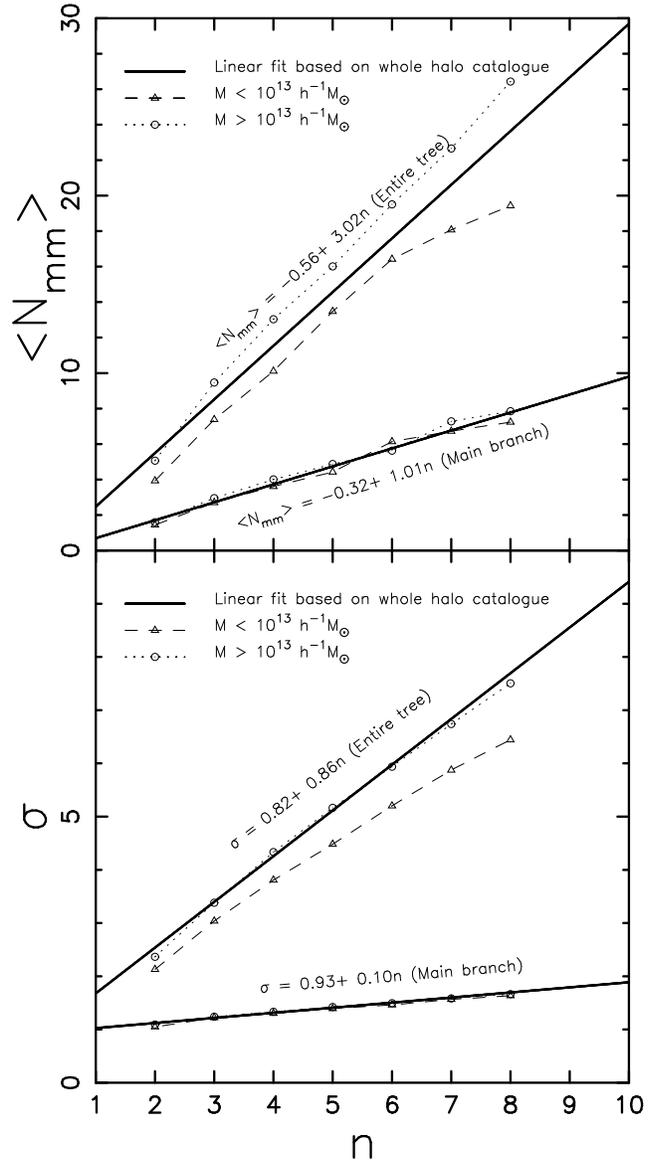}}
\caption{The median (upper panel) and dispersion (lower panel) of the 
  number  distributions of mergers with  a   mass ratio $M_1/M_2  \geq
  1/n$, as a  function of $n$. Steeper lines in each panel are the data
  from all progenitors (summing over all branches of the merger trees) 
  while flatter lines are the results from the main branch. 
  In both cases, we have divided haloes into two mass bins as indicated
  in each panel. Open triangles connected with dashed lines show the 
  results for haloes with masses $<10^{13}h^{-1}{\rm M}_\odot$,
  while open circles connected with dotted lines show the results for 
  haloes with masses $\ge 10^{13}h^{-1}{\rm M}_\odot$. 
  The  solid lines  are  the linear regressions of the data 
  drawn from the whole halo catalogue, with the slopes and zero points 
  indicated.}
\label{fig:mmfit}
\end{figure}

As pointed out above, the progenitor mass ratio used to define a major
merger  is  quite  arbritrary.    We therefore   also  investigate the
frequency of mergers with  a mass ratio larger  than $1/n$  with $n=2,
4,5,6,7,8$ (in addition to the $n=3$ discussed thus far). We find
that even with  these values of $n$ the  distributions of $N_{\rm mm}$
are still  virtually  independent  of halo mass.    
This allows  us to
consider a single $N_{\rm mm}$-distribution for  haloes of all masses. 
Fig.~\ref{fig:mmstat} plots these distributions for three different values
of  $n$ as indicated.  Each of  these distributions is reasonably well
described by  a Gaussian function (dashed curves).   Note that the use
of  a Gaussian function is  not  entirely appropriate, because $N_{\rm
  mm}$ cannot be negative. However, since  the median value of $N_{\rm
  mm}$ is, in  all cases, significantly  larger than the width  of the
distribution,  a Gaussian fit is still   appropriate.  To show how the
$N_{\rm mm}$-distribution   depends on $n$,   we plot, as in
Fig.~\ref{fig:mmfit},  the    median   and   the dispersion   of  this
distribution as functions of $n$.  As one can see, both the median and
the dispersion increase roughly  linearly with $n$,  but the slope for
the median ($\sim   1$) is much larger than   that for the  dispersion
($\sim 0.1$). Note that the results for haloes with masses 
$<10^{13}h^{-1}\Msun$ and $>10^{13}h^{-1}\Msun$ are similar, 
suggesting the distribution of the number of major mergers
is quite independent of halo mass.
  
Thus far we have only focused on the (major) merger events that merge
into  the main  branch  of the merger  tree.  For comparison, we also
consider  the merger rates  of  {\it all} progenitors, independent  of
whether they are  part of the main branch   or not. As before  we only
consider progenitors with masses in excess of one percent of the final
halo mass. The skewer lines  in Fig.~\ref{fig:mmfit} show the median  and
dispersion of  the number of  such mergers as functions  of $n$.  Here
again,  both the median  and  dispersion have roughly linear relations
with $n$. The median  number of  such major  mergers is roughly  three
times as  high as   that of major  mergers  associated  with the  main
branch, and the dispersion increases with $n$ much faster.

\begin{figure}
\centerline{\psfig{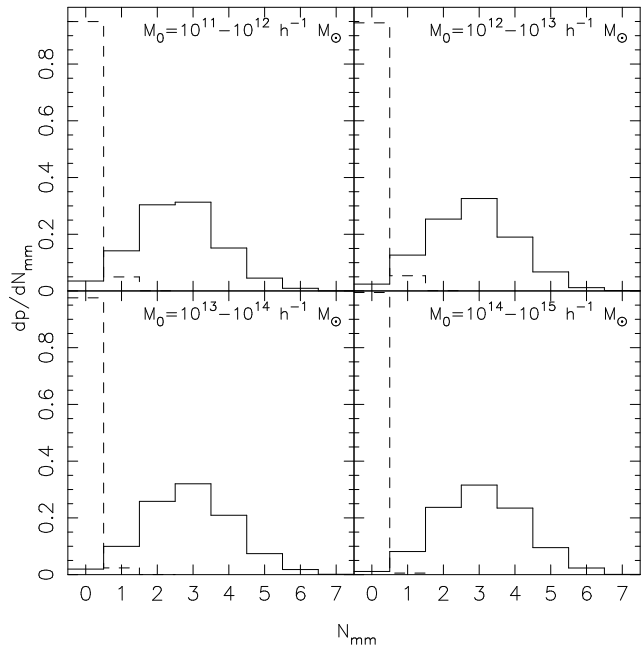}}
\caption{The probability distributions of the number of major mergers
  (those with a mass ratio larger than $1/3$) before (solid lines) and
  after (dashed lines)  $\zvcpeak$.  Note that  the vast  majority  of
  major mergers occur at $z > \zvcpeak$, demonstrating that the growth
  of the halo's virial velocity is mainly driven by major mergers.}
\label{VpkNmmBA}
\end{figure}

As mentioned  above, major mergers  are expected  to be accompanied by
rapid changes of the halo's  potential well, due  to a resulting phase
of  violent  relaxation.  To  show   such  relation  in  more  detail,
Fig.~\ref{VpkNmmBA}  shows the  distributions  of the  number of major
mergers (defined with $n=3$)  before and after the formation  redshift
$\zvcpeak$. For haloes in all mass ranges, only  a very small fraction
(less than  5\%)  experiences a major  merger  at $z<\zvcpeak$.   This
demonstrates  once again  that the growth  of  the virial  velocity is
mainly caused by major mergers. This result may have important 
implications for understanding the structure of dark matter halos.
As shown in Lu et al. (2006), if the buildup of the potential well 
associated with a dark matter halo is through major mergers, then 
the velocities of dark matter particles may be effectively 
randomized, a condition that may lead to a density profile close 
to the universal density profile observed in $N$-body simulations.   
Also, if galaxy disks are formed during a period when 
no major mergers occur, our result suggests that the potential 
wells of the halos of spiral galaxies should change little during 
disk formation.   

\section{Conclusions}
\label{sec:concl}

In the current paradigm, galaxies are thought to form in extended cold
dark  matter haloes.  A   detailed understanding of galaxy  formation,
therefore, requires a detailed  understanding of how these dark matter
haloes assemble. Halo  formation histories are typically studied using
either numerical  simulations, which are time  consuming, or using the
extended Press Schechter  formalism,  which has  been shown  to be  of
insufficient accuracy.  In this paper, we have investigated the growth
history of dark  matter haloes using  the Lagrangian perturbation code
PINOCCHIO, developed by  Monaco \etal  (2002a).  We have  demonstrated
that  the mass assembly histories (MAHs)  obtained by PINOCCHIO are in
good agreement with those obtained  using N-body simulations.  Since
PINOCCHIO  is very fast to run,  does not require any special hardware
such as supercomputers or  Beowulf clusters, and  does not require any
labor intensive  analysis, it provides a unique  and  powerful tool to
study the  statistics and assembly histories  of large samples of dark
matter haloes for different cosmologies.

Confirming  earlier results based on N-body  simulations (e.g.  W02;
Zhao \etal 2003a,b), we  find that typical MAHs  can be separated into
two phases: an early, fast accretion phase dominated by major mergers,
and  a  late, slow accretion phase   during which the   mass is mainly
accreted from minor  mergers.  However, the  MAHs of individual haloes
are complicated, and therefore difficult to parameterize uniquely by a
single parameter. We therefore defined four different formation times:
the time when  a halo acquires half of  its final mass, the time  when
the   halo's potential  well  is  established, the   time  when a halo
transits from the  fast accretion phase to  the  slow accretion phase,
and  the time when a  halo experiences its last  major merger. Using a
large  number of MAHs of  haloes spanning a  wide  range in masses, we
studied the correlations  between  these four formation  redshifts, as
well as their halo mass dependence.  Although all four formation times
are correlated, each correlation reveals a larger amount of scatter.

For all four formation redshifts, it is found that more massive haloes
assemble   later,  expressing  the hierarchical   nature  of structure
formation.  Haloes   with masses below   the characteristic non-linear
mass scale, $M^{*}$, establish their  potential wells well before they
have acquired half  of their present   day mass.  The  potential wells
associated with more massive haloes, however,  continue to deepen even
at the present  time. The time when  a halo reaches its maximum virial
velocity roughly  coincides with the time where  the MAH transits from
the fast to the slow accretion phase.

If we  define major  mergers as   those with  a progenitor mass  ratio
larger than $1/3$, then on average each halo experiences about 3 major
mergers  after its  main  progenitor has  acquired one  percent of its
present day mass. In addition, we found that the number of major 
mergers the main branch of the merging tree has experienced is 
linearly correlated with the mass ratio between the merging progenitors. 
For the whole merging tree, the number of major mergers is about 3
times that of the major mergers in the main branch. The distribution of 
the number  of major  mergers a halo  has experienced is virtually   
independent of its  mass, and the ratio between the halo  mass 
immediately after  the last major  merger and the final halo mass has a  
very broad distribution, implying that the role played by major mergers  
in building up the final halo can differ significantly from system to 
system. 

\section*{Acknowledgments}

We are grateful to Pierluigi  Monaco, Tom Theuns and Giuliano  Taffoni
for making  their wonderful code PINOCCHIO  publicly available with an
easy to understand manual, and to Xi Kang for letting us share his
EPS merging tree code. We also thank the Shanghai Supercomputer 
Center, the grants from NFSC (No. 10533030) and Shanghai Key Projects 
in Basic Research (No. 05XD14019) for the N-body simulations 
used in this paper. HJM would like to acknowledge the support of 
NSF AST-0607535, NASA AISR-126270 and NSF IIS-0611948.
FvdB acknowledges useful and lively  discussions with Risa Wechsler  
during  an early phase of this project.

\label{lastpage}

\end{document}

%% file: macros_yun.tex
\newcommand{\dd}{{\rm d}}
\newcommand{\dS}{\Delta S}
\newcommand{\dW}{\Delta \omega}
\newcommand{\deltac}{\Delta_{\rm vir}}

\newcommand{\Mh}{M_{\rm vir}}
\newcommand{\Rh}{R_{\rm vir}}
\newcommand{\Vh}{V_{\rm vir}}

\newcommand{\zhalf}{z_{\rm half}}
\newcommand{\zvczero}{z_{\rm vvir}}
\newcommand{\zvcpeak}{z_{\rm vmax}}
\newcommand{\zlmm}{z_{\rm lmm}}

\newcommand{\bfx}{\textbf{x}}
\newcommand{\bfq}{\textbf{q}}
\newcommand{\bfS}{\textbf{S}}

\newcommand{\etal}{{et al.~}}

\newcommand{\gta}{\ga}

\newcommand{\Mpc}{\>{\rm Mpc}}

\newcommand{\Msun}{\>{\rm M_{\odot}}}
\newcommand{\msun}{\>{\rm M_{\odot}}}

\newcommand{\beq}{\begin{equation}}
\newcommand{\eeq}{\end{equation}}

\newcommand{\mpch}{\>h^{-1}{\rm {Mpc}}}

\newcommand{\apj}{ApJ}

\newcommand{\mnras}{MNRAS}
\newcommand{\aap}{A\&A}

\newdimen\hssize
\newdimen\hdsize 
\def\ltsima{$\; \buildrel < \over \sim \;$}
\def\lsim{\lower.5ex\hbox{\ltsima}}
\def\gtsima{$\; \buildrel > \over \sim \;$}
\def\gsim{\lower.5ex\hbox{\gtsima}}